# Time Interval Distribution of Earthquakes

Sumiyoshi Abe[1] and Norikazu Suzuki[2]

[1]Institute of Physics, University of Tsukuba, Ibaraki 305-8571, Japan

[2]College of Science and Technology, Nihon University, Chiba 274-8501, Japan

Although seismicity is characterized by extraordinarily rich phenomenology and complexity, which make its coherent explanation and prediction difficult, some empirical laws are known to hold, which are remarkably simple. Examples are the Omori law *(1)* for temporal pattern for aftershocks and the Gutenberg-Richter law *(2)* for frequency and magnitude. These are power laws and represent scale free natures of earthquakes.

One of the extreme goals of seismology is to predict when the next significant shock will come after some earlier significant shock. For this, it is of crucial importance to elucidate the statistical properties of observed random nonstationary time series of magnitude.

In this article, we report the discovery of a new scaling law concerning the distribution of time intervals between significant earthquakes. In particular, we show that this distribution is of the Zipf-Mandelbrot type *(3)*.

We have analyzed a set of observed time series data of magnitude made available by the Southern California Earthquake Data Center (http://www.scecdc.scec.org/catalogs.html). To measure time interval between two significant (or, bodily sensible) shocks, we have set up a threshold value of magnitude, which is defined here by the mean value plus half of the standard deviation. In fact, the result turned out be insensitive to the definition of the threshold. We have measured the interval between two successive values of magnitude above the threshold. This interval is referred to here as the "calm time interval", $\tau$. A collection of calm time intervals thus obtained gives a histogram, from which the cumulative probability, $P(>\tau)$, could be calculated.

In Fig. 1, we present the log-log plot of an observed cumulative distribution associated with the statistical frequency of the calm time interval represented by the blue dots, whereas the inside box is drawn on the "semi-*q*-log scale". The red line shows the Zipf-Mandelbrot type distribution, $P(>\tau) = e_q(-\tau/\tau_0)$, where $e_q(x)$ stands for the so-called *q-exponential function* defined by $e_q(x) = [1+(1-q)x]^{1/(1-q)}$ if $1+(1-q)x \geq 0$ and $e_q(x) = 0$ if $1+(1-q)x < 0$. The *q-logarithmic function* used in the inside box is the inverse function of the *q*-exponential function and is explicitly given by $\ln_q(x) = (x^{1-q}-1)/(1-q)$. The ordinary Zipf-Mandelbrot distribution corresponds to the exponent of $q > 1$, i.e., a power-law distribution with a long tail, as in Fig. 1. Therefore, in addition to the Omori and Gutenberg-Richter laws, a new scale free nature of earthquakes has been identified.

Following analysis of other data taken from the same site, we have ascertained that the trend presented here is universal. We have also recognized that remarkably the seismic time series always follows the Zipf-Mandelbrot distribution with $q > 1$.

In conclusion, we have discovered a new scaling law for calm time intervals between significant earthquakes. We have observed that the seismic time series undergoes a series of states, which are all described by the Zipf-Mandelbrot distributions with various values of the exponent, $q$. It may also be worth mentioning that there exists a statistical mechanical theory for such distributions. The generalized entropy indexed by $q$ proposed in *(4)* is known to be maximized by them under appropriate constraints on the averages of the physical quantities (e.g., the calm time interval) to be measured. An important point is that this fact enables us to develop a thermodynamic approach to seismology and should provide new insights into the study of earthquakes.

# Figure Caption

Fig. 1 The log-log and the semi-$q$-log plots of an observed cumulative distribution associated with the frequency of the calm time interval. The dots and the solid lines represent the data from the Southern California Earthquake Data Center and the Zipf-Mandelbrot distribution, respectively. The data were taken between 08:08:41.26 on 19 October, 1999 and 09:13:19.27 on 1 November, 1999. The number of data values is 3056. The threshold of magnitude is 2.4. The parameters of the distribution are $q = 1.36$ and $\tau_0 = 1.0 \times 10^3 \, \text{s}$. The straight line on the semi-$q$-log scale corresponds to the Zipf-Mandelbrot distribution, with the value of correlation coefficient $\rho = -0.997$.

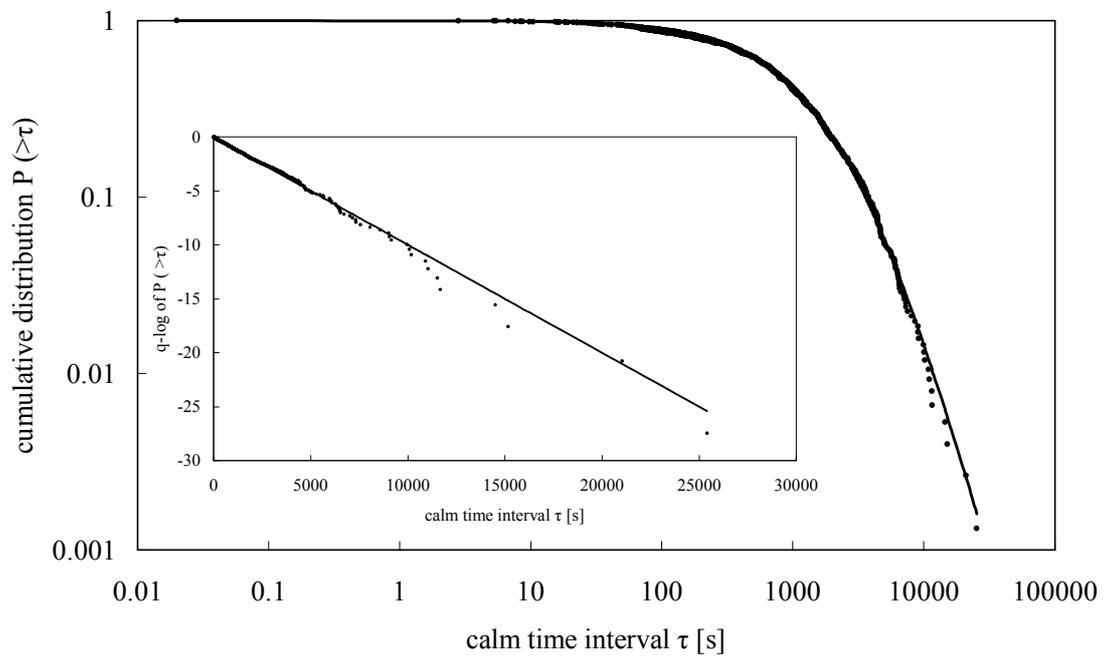

**Fig. 1**